\documentclass[aps,preprint,floatfix,altaffilletter]{revtex4}
\usepackage{dcolumn}
\usepackage{color}
\usepackage{amsmath}
\usepackage{amssymb}

\usepackage[dvips]{graphicx}

\newcommand{\be}{\begin{equation}}
\newcommand{\ee}{\end{equation}}
\newcommand{\bea}{\begin{eqnarray}}
\newcommand{\eea}{\end{eqnarray}}

\newcounter{Fig}

\begin{document}


\title{Magnetized Epsilon-Near-Zero (ENZ) Structures:  Hall Opacity, Hall Transparency, and One-Way Photonic Surface States}

\author{Arthur R. Davoyan}
\email{davoyan@seas.upenn.edu}

\author{Nader Engheta}
\email{engheta@ee.upenn.edu}

\affiliation{Department of Electrical and Systems Engineering, University of Pennsylvania, Philadelphia,
Pennsylvania 19104, USA }

\date{\today}


\begin{abstract}
We study propagation of transverse-magnetic (TM) electromagnetic waves in the bulk and at the surface of magnetized epsilon-near-zero (ENZ) medium in a Voigt configuration. We reveal that in a certain range of material parameters novel regimes of wave propagation emerge: we show that the transparency of the medium can be altered with the magnetization leading either to magnetically induced Hall opacity or Hall transparency of the ENZ. In our theoretical study, we demonstrate that surface waves at the interface between either a transparent or an opaque Hall medium and a homogeneous medium may, under certain conditions, be predominantly one-way. Moreover, we predict that one-way photonic surface states may exist at the interface of an opaque Hall ENZ and a regular metal, giving rise to a possibility for backscattering immune wave propagation and isolation.
\end{abstract}

\maketitle

Electromagnetic properties of any media are fully described by its electric and magnetic responses, which in the simplest case are given by scalar electric permittivity ($\varepsilon$) and magnetic permeability ($\mu$)~\cite{Landau,Born}. Depending on the signs of the real parts of the permittivity and permeability naturally occurring materials can be classified into three large groups: opaque metals ($\varepsilon<0$, $\mu>0$), opaque magnetics ($\mu<0$, $\varepsilon>0$) and transparent dielectrics ($\mu>0$, $\varepsilon>0$). Recent progress with metamaterials -- artificially engineered structures for the light-matter interaction -- allows achieving simultaneously negative permittivity and permeability ($\varepsilon<0$, $\mu<0$)~\cite{Pendry,Shalaev,Zhang}, as well as designing structures with epsilon-near-zero ($\varepsilon\simeq0$, $\mu\simeq1$)~\cite{ENZ,Engheta_ENZ}, and epsilon- and mu-near-zero ($\varepsilon\simeq0$ and $\mu\simeq0$)~\cite{EMNZ} electromagnetic responses. The ability to design and utilize materials with properties not readily available in nature may lead to intriguing physics and serve as a platform for a variety of applications practically in all areas of electrodynamics.

Mixing electric and magnetic responses of the medium, for example, by utilizing magneto-active or bi-anisotropic effects might give additional degrees of freedom for tailoring the light propagation~\cite{Landau}. Most importantly, magnetized structures with magneto-active response, in a sharp contrast to conventional scalar materials, might break the time-reversal symmetry in the electromagnetic phenomena~\cite{Camley}. Furthermore, it was suggested that waves propagating at the interface of magnetized medium can be nonreciprocal, so that waves propagating in positive ($+x$) and negative ($-x$) directions have different dispersion properties~\cite{Camley,nonreciprocal,Kushwaha,Davis,Such_v1,Such_v2}. This feature has been utilized for optical isolation~\cite{Fan_isolat,Khanikaev} and for one-way backscattering immune surface wave propagation~\cite{Haldane,Fan_oneway,Soljacic_oneway,Khanikaev_nat}. Employing metamaterial principles, it might be possible to deliberately design the magneto-active response of the media, i.e. achieve desired values for the corresponding permittivity (or permeability) tensor~\cite{Davoyan}.

In this Letter, we study phenomenologically the electromagnetic wave propagation in a magneto-active medium described by an effective permittivity tensor, with particular attention towards materials with near-zero relative permittivity (i.e., epsilon-near-zero (ENZ) media). Using analytical techniques and numerical simulation, we demonstrate that in a certain range of parameters of the magnetized material conceptually new types of electromagnetic response emerge, which we classify as Hall opacity and Hall transparency. We study propagation of the surface waves at the interface of such magnetized medium and reveal nonreciprocal and one-way regimes of wave propagation. We show that surface waves at the interface of the opaque and transparent Hall medium are predominantly one-way in the entire range of possible material parameters. Furthermore we demonstrate that one-way surface waves can exist at the boundary of two opaque media. Such solutions can be utilized for the scattering immune wave guiding and wave isolation.

\begin{figure}[t]
  \begin{center}
      \includegraphics[width=1\columnwidth]{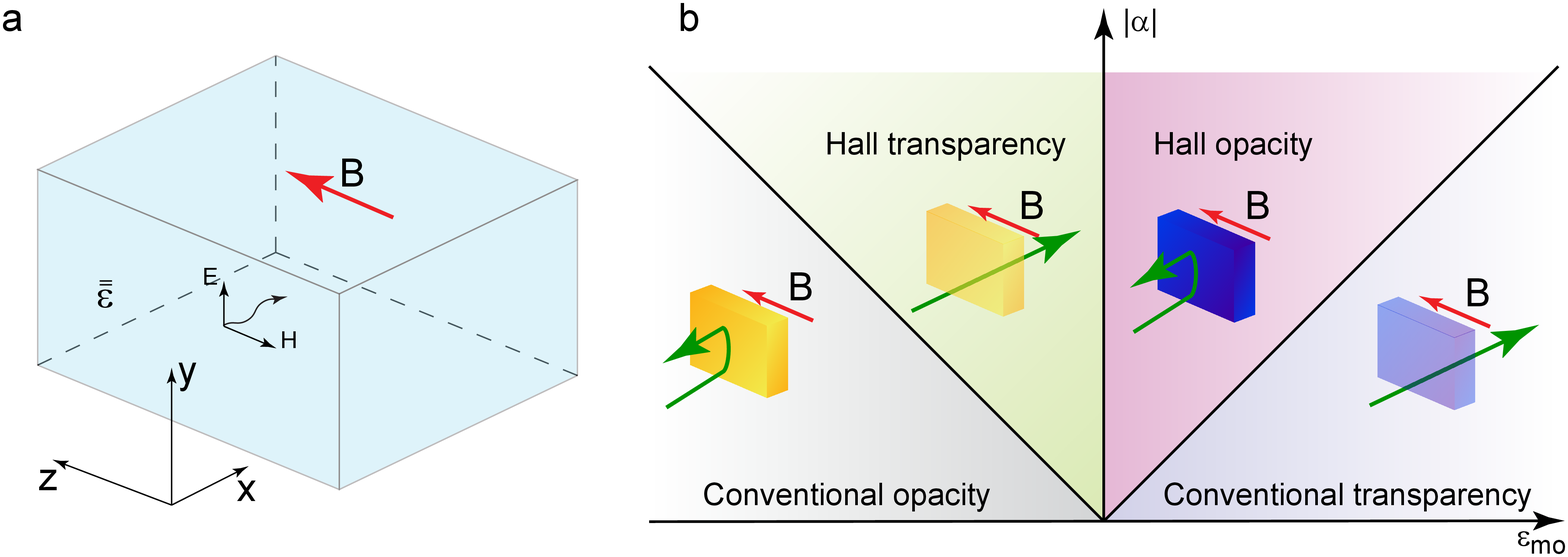}
      \caption{(color online) (a) Geometry of the problem of the bulk magnetized medium including the case of epsilon-near-zero medium (ENZ) material. Direction of wave propagation is perpendicular to the magnetization direction (Voigt configuration). (b) Categorization of materials depending on the values of diagonal and off-diagonal components of relative permittivity tensor.} \label{geometry}
  \end{center}
\end{figure}

We begin with the study of the wave propagation in a bulk magnetized medium, as shown schematically in Fig.~\ref{geometry}(a). In our analysis we assume that the magnetization of the medium influences only the electric permittivity, whereas magnetic permeability stays unperturbed (i.e. $\mu\simeq1$). In this case the dielectric response of the medium is described by an Hermitian antisymmetric relative permittivity tensor~\cite{Landau}:
 \begin{equation}\label{tensor}
  \bar{\bar{\varepsilon}}= \left( \begin{array}{ccc}
   \varepsilon_{mo} & i\alpha & 0 \\
   -i\alpha & \varepsilon_{mo} & 0 \\
   0 & 0 & \varepsilon_\bot \end{array} \right),
\end{equation}
where $\varepsilon_{mo}$ and $\varepsilon_\bot$ are diagonal components of relative dielectric permittivity, $\alpha$ is the off-diagonal component of the permittivity tensor responsible for the ``strength'' of magneto-optical activity of the media. For clarity of our analysis and without loss of generality we shall assume further that $\alpha>0$. Moreover, for the sake of ease of discussion and to highlight the main physical concepts, we assume that the materials are all lossless, and therefore the parameters $\varepsilon_{mo}$, $\varepsilon_{\bot}$, and $\alpha$ are real quantities. We also consider that the wave propagation is in a direction perpendicular to the magnetization direction, i.e. the Voigt configuration, see Fig.1(a). (The wave propagation along the magnetization direction, i.e. in the Faraday geometry, is quite intuitive and well studied in the literature~\cite{Landau,Davoyan}). According to the relative permittivity tensor (\ref{tensor}) TM $(E_x,E_y,0)$ and TE $(0,0,E_z)$ waves in such a geometry stay uncoupled, and hence can be studied independently. We note that due to the symmetry of the Maxwell equations the effects predicted in this Letter can be easily extended via duality to the materials with the permeability tensor (ferromagnetic response) as well~\cite{Camley,Kushwaha}.

The propagation of TM waves in such a medium is described by the following dispersion relation~\cite{Landau}:
\begin{equation}
  k_x^2+k_y^2=\left(\frac{2\pi}{\lambda}\right)^2 \frac{\varepsilon_{mo}^2-\alpha^2}{\varepsilon_{mo}}, \label{dispersion_bulk}
\end{equation}
where $k_x$ and $k_y$ are the wavevector components along the $x$ and $y$ directions, respectively, and $\lambda$ is the free-space wavelength.

First, it should be noted that despite the anisotropic nature of the permittivity tensor the wave propagation in the $(x-y)$ plane is isotropic, and is defined by an effective permittivity $\varepsilon_{eff}=(\varepsilon_{mo}^2-\alpha^2)/\varepsilon_{mo}$. Similarly to conventional materials with scalar dielectric response, this TM wave propagation in the magnetized medium depends on the sign of $\varepsilon_{eff}$. In particular, the medium is opaque for TM waves when $\varepsilon_{eff}<0$ and is transparent when $\varepsilon_{eff}>0$. However, the sign of the effective permittivity depends on the relation between $\varepsilon_{mo}$ and $\alpha$. In Fig.1(b) we map the regimes of TM wave propagation in the $(\alpha,\varepsilon_{mo})$ plane in an analogy with the ($\varepsilon-\mu$) material parameters map. Note that for $\alpha=0$, i.e. nonmagnetized medium, the material properties depend on the sign of the $\varepsilon_{mo}$ only, and can be classified as a usual metal and a usual dielectric, as expected. For nonzero $\alpha$ we see an emergence of two novel types of material response, which we classify as opaque Hall medium and transparent Hall medium (Fig.~\ref{geometry}(b)). In particular, even for negative $\varepsilon_{mo}$, i.e. initially metallic state, the magnetization may induce transparency in the medium when $-|\alpha|<\varepsilon_{mo}<0$. In the range of parameters $0<\varepsilon_{mo}<|\alpha|$ we notice an opposite behavior -- the medium becomes opaque for TM waves. It should be noted here that many naturally occurring materials have $|\alpha|\ll\varepsilon_{mo}$, and the behavior predicted here is typically not observed. However, employing metamaterial concepts it is possible to design structures with epsilon-near zero response, i.e. $|\varepsilon_{mo}|\rightarrow 0$, so that the condition $|\alpha/\varepsilon_{mo}|>1$ can be achieved~\cite{Engheta_ENZ,Davoyan}.

\begin{figure}[t]
  \begin{center}
      \includegraphics[width=1\columnwidth]{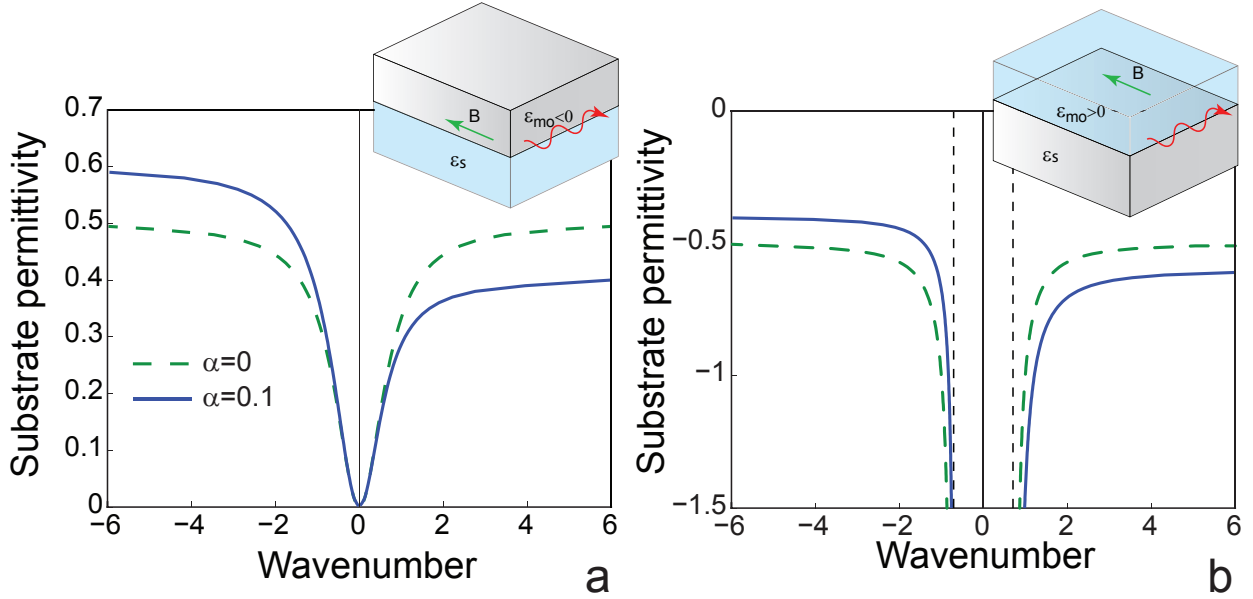}
      \caption{(color online) Dispersions of right- and left-going surface waves at the interface of (a) magnetized metal ($\varepsilon_{mo}=-0.5$) and a magnetized dielectric ($\varepsilon_{mo}=+0.5$) for two different values of magnetization $\alpha$. Vertical dashed lines in panel (b) show the light-lines ($\beta=\sqrt{\varepsilon_{eff}}$) in the magnetized medium ($\alpha=0.1$). Insets show the geometries of the structures.}\label{fig_disp_surf}
  \end{center}
\end{figure}

Despite the medium being magnetized, the analysis of the dispersion relation (\ref{dispersion_bulk}) shows that the TM wave propagation in the ($x-y$) plane is reciprocal, i.e. this wave propagation is direction invariant. By reducing the dimensionality of the system, i.e. by considering surface wave propagation at the interface between such a magnetized medium and a conventional material, it is possible to break the reciprocity in the wave propagation phenomenon~\cite{nonreciprocal,Camley}. In order to get a better understanding of the surface-induced nonreciprocity we study propagation of surface waves at the interface of a magnetized medium with a neighboring material (denoted as ``substrate" in Fig.~\ref{fig_disp_surf}). The general dispersion relation for the surface TM waves in this case can be presented as follows:
\begin{equation} \label{dispersion_surf}
   \frac{\beta}{\varepsilon_{eff}}\frac{\alpha}{\varepsilon_{mo}}-\frac{\sqrt{\beta^2-\varepsilon_{eff}}}{\varepsilon_{eff}} = \frac{\sqrt{\beta^2-\varepsilon_s}}{\varepsilon_s}
\end{equation}
where $\beta$ is the normalized surface wave wavenumber, i.e. $\beta=k/(2\pi/\lambda)$, where $k$ is the wavevector of the surface wave, and $\varepsilon_s$ is the permittivity of the neighboring (substrate) medium, which can be either negative (metal) or positive (dielectric). Clearly, for a given direction of magnetization, i.e. given sign of $\alpha$, the waves with $\beta>0$ and $\beta<0$ have different properties. For the parametric analysis of this dispersion relation here, it is convenient to search for the solutions of the dispersion relation, i.e., $\beta$, as the permittivity of the substrate is varied, when given values of magneto-optical permittivity tensor elements are chosen. Figures~\ref{fig_disp_surf}(a,b) show typical dispersions for surface waves at the interface of the magnetized metal (i.e., $\varepsilon_{mo}<0$) (Fig.~\ref{fig_disp_surf}(a)) and magnetized dielectric (i.e., $\varepsilon_{mo}>0$ (Fig.~\ref{fig_disp_surf}(b)) with the substrate when $|\varepsilon_{mo}|>\alpha$. When the magnetization is absent, i.e. $\alpha=0$ the dispersion curves are symmetric and correspond to the well-known dispersion of the surface plasmon-polaritons at a single metal-dielectric interface~\cite{Maier_rev}, see dashed curves in Figs.~\ref{fig_disp_surf}(a,b). Near the surface plasmon resonance corresponding to the condition $|\varepsilon_{mo}|=|\varepsilon_s|$ the wavelength of surface wave becomes infinitesimally small, i.e. $|\beta|\rightarrow\infty$. When the magnetization is introduced in the system, the symmetry between surface waves with positive and negative $\beta$ is broken. Furthermore, the surface plasmon resonance condition differs, i.e., ``splits" for $\beta>0$ and $\beta<0$. In particular, for $\beta\rightarrow+\infty$ the resonance occurs at $\varepsilon_s = -\varepsilon_{mo}-\alpha$, whereas for opposite phase progression direction, i.e. $\beta<0$ the resonance condition is at $\varepsilon_s = -\varepsilon_{mo}+\alpha$. Such a splitting between the resonances implies that in the range of parameters $-\varepsilon_{mo}-\alpha<\varepsilon_s<-\varepsilon_{mo}+\alpha$ only one-way surface states exist (in this case with $\beta<0$). This effect of one-way surface wave propagation has been employed recently for the design of unidirectional scattering immune waveguides~\cite{Fan_oneway} and for one-way loads and antennas~\cite{load}.

\begin{figure}[h]
  \begin{center}
      \includegraphics[width=1\columnwidth]{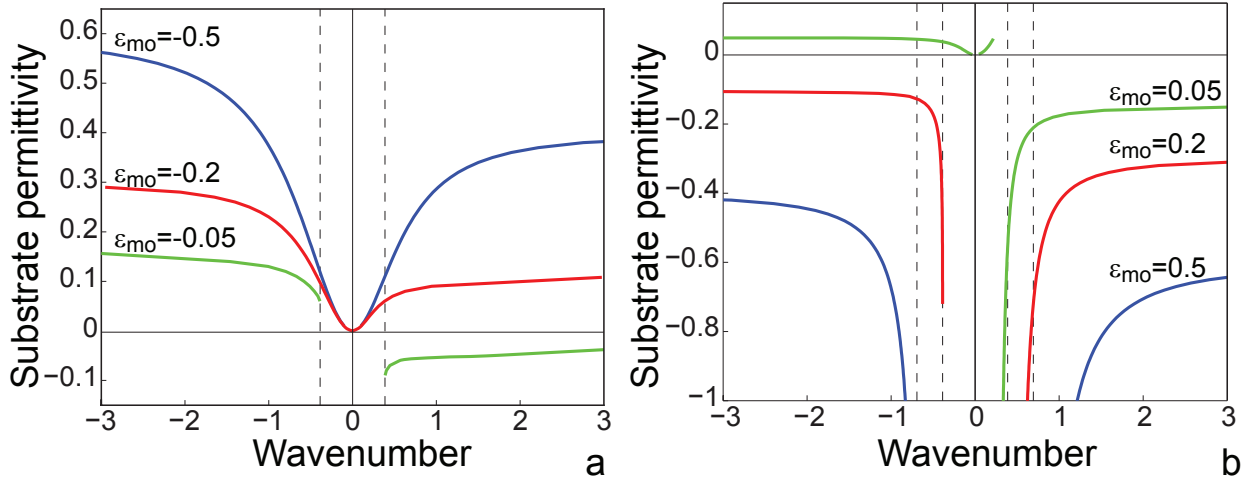}
      \caption{(color online) Dispersion of the surface waves for the cases of (a) magnetized metal, and (b) magnetized dielectric, for different values of the diagonal component of the permittivity tensor, and $\alpha=0.1$. Vertical dashed lines correspond to the light-lines in the magnetized medium.}\label{fig_disp_trans}
  \end{center}
\end{figure}

When approaching the regimes of Hall opacity and Hall transparency, i.e. $|\varepsilon_{mo}|\rightarrow \alpha$ the dynamics of the surface waves with both positive ($\beta>0$) and negative ($\beta<0$) directions of phase progression changes dramatically. In Figs.~\ref{fig_disp_trans}(a,b) we plot the dispersions of waves propagating at the surface of a magnetized metal (Fig.~\ref{fig_disp_trans}(a)) and magnetized dielectric (Fig.~\ref{fig_disp_trans}(b)) for a given value of magneto-optical activity $\alpha=0.1$ and for different values of $|\varepsilon_{mo}|$. For a magnetized metal we observe that with the decrease of the ratio $|\varepsilon_{mo}|/\alpha$ the required substrate relative permittivity for the surface plasmon resonance, i.e. $\varepsilon_s = -\varepsilon_{mo}\pm\alpha$, also decreases. When $|\varepsilon_{mo}|\rightarrow \alpha$ the dispersion curve for positive $\beta$ diverges, so that bound surface waves exist only for $\beta<0$ in entire range of substrate permittivity. The condition $|\varepsilon_{mo}| = \alpha$ corresponds to the transition from an opaque metallic phase to the Hall transparency phase, see Fig.~\ref{geometry}(b), and is followed by the bifurcation in the dispersion curves. In particular, light lines ($\beta=\pm\sqrt{\varepsilon_{eff}}$) appear, which bound the surface wave dispersions from lower $|\beta|$ limit. (Note that we consider only the localized bound (guided) solutions, and do not take into account solutions corresponding to the leaky waves). We notice that bound modes with $\beta<0$ exist only in the range of positive substrate permittivities $\varepsilon_s>0$, implying that surface states exist on the interface between two transparent media with positive effective bulk permittivities. On the other hand, surface waves with $\beta>0$ exist only at the interface with negative permittivity substrate. Noticeably the interface of the transparent Hall medium supports only one-way surface waves.

The surface wave propagation on the surface of the magnetized dielectric, i.e. $\varepsilon_{mo}>0$, is shown in Fig.~\ref{fig_disp_trans}(b). First, we notice that the dispersion curve for the left-going wave even for the case $\varepsilon_{mo}\gg\alpha$ (see also Fig.~\ref{fig_disp_surf}(b)) is ``shifted" towards the light-line. Searching for the solutions of the dispersion equation (\ref{dispersion_surf}) in the limit $\varepsilon_s\rightarrow -\infty$ it is possible to show that only waves with $\beta>0$ exist in this limit. The absence of solutions for $\beta<0$ implies that at some value of $\varepsilon_s$ in the range of parameters $-\infty<\varepsilon_s<-(\varepsilon_{mo}-\alpha)$ the waves with $\beta<0$ become leaky, i.e. they cross the light-line ($\beta=-\sqrt{\varepsilon_{eff}}$) in the magnetized medium as a specific $\varepsilon_s$ is given below~\cite{Kushwaha,Davis}. Hence, below this regime only ``one-way" bound surface waves with $\beta>0$ exist. The detailed discussion of this regime of wave propagation, its feasibility with realistic materials and its applications are outside the scope of this Letter, and will be reported in detail in a future publication.  The critical value of the substrate permittivity for the light-line crossing is given by $\varepsilon_s^c=-0.5\varepsilon_{eff}(1+\sqrt{4+\varepsilon_{mo}^2/\alpha^2})\varepsilon_{mo}^2/\alpha^2$. Clearly, for $\varepsilon_{mo}\gg\alpha$ and $\alpha\rightarrow 0$ this regime is possible only for $\varepsilon_s\ll-1$. Hence, two regimes of one-way surface wave propagation at the interface between a magnetized dielectric and a metallic substrate exist: one-way surface waves with positive phase velocity ($\beta>0$) exist for the substrate relative permittivity values below the light-line crossing, i.e. for $\varepsilon_s<\varepsilon_s^c$, and one-way surface waves with negative phase velocity ($\beta<0$) exist in the range $-(\varepsilon_{mo}+\alpha)<\varepsilon_s<-(\varepsilon_{mo}-\alpha)$. The latter is clearly seen for $\varepsilon_{mo}$ approaching $\alpha$, see Fig.~\ref{fig_disp_trans}(b). With further decrease of the ratio $\varepsilon_{mo}/\alpha$ the range of existence of $\beta<0$ solutions shrinks and degenerates, so only positive $\beta$ surface waves exist. At the values of $\varepsilon_{mo} = \alpha$ transition from transparent dielectric to the opaque Hall medium occurs, which also is followed with the corresponding bifurcation in the dispersion curves. In particular, negative phase velocity waves, i.e. with $\beta<0$, emerge for positive values of the substrate permittivity.
\begin{figure}[h]
  \begin{center}
      \includegraphics[width=1\columnwidth]{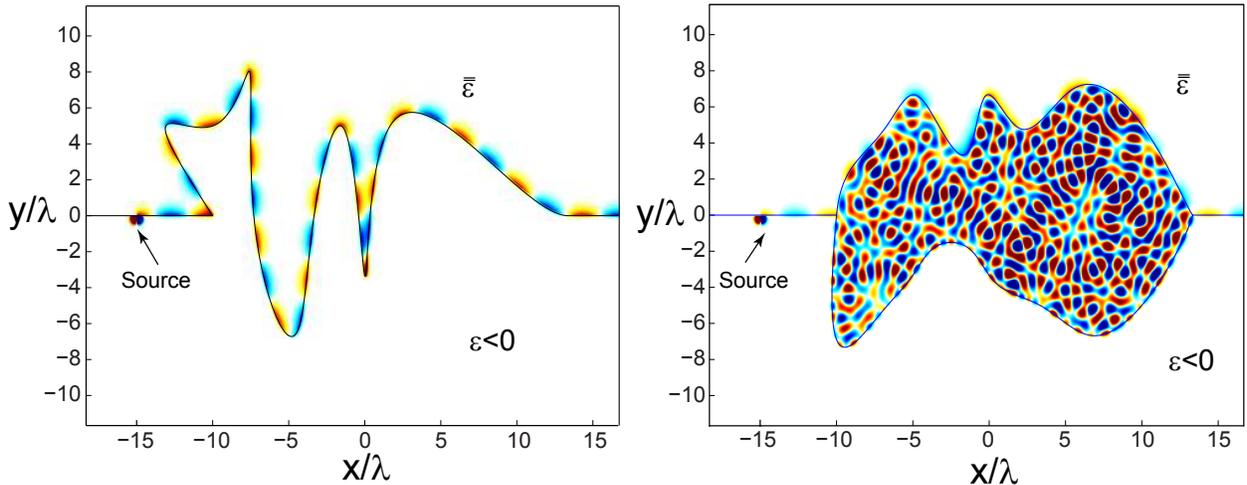}
      \caption{(color online) Simulation results for the one-way surface wave propagation along the interface of the metal and opaque Hall medium: (a) for a ``zig-zag" interface and (b) for an interface with an air cavity. Here $\varepsilon_{mo}=0.03$, $\alpha=-0.1$, $\varepsilon_s = -5$, and $\lambda$ is the free-space wavelength. Propagation into the bulk medium is forbidden, and the one-way surface state provides topologically scattering-immune system.}\label{field}
  \end{center}
\end{figure}

Interestingly, one-way surface waves with $\beta>0$ exist at the interface between two opaque media, i.e. on the interface between opaque Hall medium and a regular metal, see curve for $\varepsilon_{mo}=0.05$ in Fig.~\ref{fig_disp_trans}(b). This case is of a particular interest since one-way surface states are the only allowed propagating solutions in the system. Hence, such a wave would be topologically protected from backscattering on any kind of defects or any abrupt changes of the interface. Figure~\ref{field} shows the propagation of such a surface wave along a ``zig-zag" shaped interface. Clearly, no backscattering and no radiation into the ambient media exist. The wave follows any arbitrary shape of the interface. Similarly an air cavity placed on the interface between the two media does not influence the light flow, see Fig.~\ref{field}(b).

We note that our analysis of the power flux carried by the surface waves (i.e. $S_x  = \frac{\sqrt{\mu_0/\varepsilon_0}}{4\omega/c}\frac{1}{\varepsilon_{eff}}\left[\left(\frac{-\alpha}{\varepsilon_{mo}}+\frac{\beta}{\kappa_{eff}}\right)+\frac{\varepsilon_{eff}}{\varepsilon_s}\frac{\beta}{\kappa_s}\right]$, where $\kappa_{eff} = \sqrt{\beta^2-\varepsilon_{eff}}$ and $\kappa_s =\sqrt{\beta^2-\varepsilon_s}$) shows that the phase progression and power transfer are in the same direction, i.e. no backward modes with $\beta S_z <0$ exist in such a system for any values of the magneto-optical material and the substrate.

Finally, we note that the dispersion equation (\ref{dispersion_surf}) also allows solutions for surface waves propagating in $+x$ direction at the interface between the magnetized dielectric and a conventional dielectric with an arbitrary set of values for $\varepsilon_{mo}$ and $\alpha$. These solutions exist in the limit $\varepsilon_{mo}-\varepsilon_s\rightarrow+0$, and are weakly confined to the interface, therefore, they are not of interest for the purpose of any potential applications. It is worth pointing out that some of the features discussed above may be affected by material losses. Here we have been interested in exploring these phenomena in the limit of low (or zero) loss.

In conclusion, we have studied wave propagation in a bulk and at the surface of the magnetized medium, and revealed that for epsilon-near-zero regime the properties of wave propagation change significantly when magnetized. In particular, we have demonstrated that at the interface between a magnetized ENZ medium and a substrate predominantly one-way surface waves exist. We showed that one way surface waves exist at the interface of two opaque media, and revealed novel regimes of back-scattering immune wave propagation.

 This work is supported in part by the National Science Foundation (NSF), Materials Research Science and Engineering Center (MRSEC) Award No. DMR11-20901.

\end{document}